\documentclass[preprint2]{aastex}

\newcommand{\myemail}{wnp@mpe.mpg.de}

\def\cm-2{cm$^{-2}$}

\def\ein{{\it Einstein}}
\def\asca{{ASCA}}
\def\chandra{{\it Chandra}}

\def\xmm{{XMM-{\it Newton}}}
\def\ro{{ROSAT}}
\def\exo{{EXOSAT}}
\def\n253{\object{NGC~253}}
\def\me31{\object{M~31}}
\def\m33{\object{M~33}}
\def\mx7{\object{M~33~X$-$7}}
\def\x7{X$-$7}

\newcommand{\ergcm}[1]{$\times 10^{#1}$ \hbox{erg cm$^{-2}$ s$^{-1}$}}

\newcommand{\ergs}[1]{$\times 10^{#1}$ \hbox{erg s$^{-1}$}}
\newcommand{\oergs}[1]{$10^{#1}$ erg s$^{-1}$}
\newcommand{\hcm}[1]{$\times 10^{#1}$ cm$^{-2}$}
\newcommand{\ohcm}[1]{$10^{#1}$ cm$^{-2}$}
\newcommand{\expo}[1]{$\times 10^{#1}$}

\newcommand{\ct}{ct s$^{-1}$}

\newcommand{\EXPN}[2]{\mbox{$#1\times 10^{#2}$}}

\shorttitle{ChASeM33 observations of M\,33~X--7}
\shortauthors{Pietsch et al.}

\begin{document}


\title{M\,33~X--7: ChASeM33 reveals the first eclipsing black hole X-ray
binary}


\author{W. Pietsch and F. Haberl}
\affil{Max-Planck-Institut f\"ur extraterrestrische Physik, Giessenbachstrasse,
       85741 Garching, Germany}
\email{\myemail}

\author{M. Sasaki, T.J. Gaetz, P.P. Plucinsky}
\affil{Harvard-Smithsonian Center for Astrophysics, 60 Garden Street, Cambridge,
MA 02138, USA}

\and

\author{P. Ghavamian}
\affil{Department of Physics and Astronomy, John Hopkins University, 3400 North
Charles Street, Baltimore, MD 21218-2686, USA}

\and

\author{K.S. Long}
\affil{Space Telescope Science Institute, 3700 San Martin Drive, Baltimore, 
       MD 21218, USA}

\and

\author{T.G. Pannuti}
\affil{Spitzer Science Center, Jet Propulsion Laboratory/California Institute of Technology, Mailstop 220-6,
Pasadena, CA 91125, USA}

\begin{abstract}
The first observations conducted as part of the \chandra\ ACIS survey of 
\m33\ (ChASeM33) sampled the 
eclipsing X-ray binary \mx7\ over a large part of the 3.45 d orbital period and
have resolved eclipse ingress and egress for the first time. The occurrence of the 
X-ray eclipse allows us to determine an improved ephemeris of mid-eclipse and 
binary period as
HJD~$(245\,3639.119\pm0.005)\pm N\times(3.453014\pm0.000020)$ and constrain the 
eclipse half angle to $26.5\degr\pm1.1\degr$. There are indications for a
shortening of the orbital period.  The X-ray spectrum is best described by a disk
blackbody spectrum  typical for black hole X-ray binaries in the Galaxy.
We find a flat power density spectrum and no significant regular pulsations 
were found in the frequency range of 10$^{-4}$--0.15~Hz. 
HST WFPC2 images resolve the optical counterpart, which can be
identified as an O6III star with the help of
extinction and colour corrections derived from the X-ray absorption. 
Based on the optical light curve, the mass of the
compact object in the system most likely exceeds $9 M_{\sun}$.
This mass, the shape of the X-ray spectrum and the short term X-ray time 
variability identify 
\mx7\ as the first eclipsing black hole high mass X-ray binary.
\end{abstract}


\section{Introduction}
\mx7\ (hereafter \x7) was detected as a variable source by the \ein\
observatory with a maximum luminosity in the 0.15--4.5~keV band 
(assuming an absorption column of \ohcm{21}) that 
exceeds \oergs{38} 
\citep[][]{1981ApJ...246L..61L,1983ApJ...275..571M}. 
The source stayed active in all following observations. 
Its variability was explained by 
an eclipsing X-ray binary (XRB) with an orbital period of 1.7~d and an eclipse duration
of $\sim$0.4~d \citep{1989ApJ...336..140P,1993ApJ...418L..67S,1994ApJ...426L..55S}. 
Based on \ro\ and \asca\ data
\citep[][hereafter DCL99]{1997AJ....113..618L,1999MNRAS.302..731D}, the orbital
period was found to be twice as long. DCL99 described the shape of the eclipse
by a slow ingress 
($\Delta \Phi_{\rm ingress} = 0.10\pm0.05$), 
an eclipse duration of  $\Delta \Phi_{\rm eclipse} = 0.20\pm0.03$ and a fast
eclipse egress ($\Delta \Phi_{\rm egress} = 0.01\pm0.01$) with an ephemeris for
the mid-eclipse time of HJD~244\,8631.5$\pm$0.1 + N$\times$(3.4535$\pm$0.0005).
DCL99 also report 3$\sigma$ evidence for a 0.31~s pulse period. They come to
that conclusion by splitting PSPC and HRI data of 500~s intervals of continuous
data where \x7\ was positively detected and calculating the summed Rayleigh power
spectrum. The linearly binned power spectrum showed an significant excess (at
99.9 per cent confidence) at the proposed period when compared to simulated data
assuming a Poisson distribution with the \x7\ mean flux. The signal is broader
than what would be expected from a simple sinusoidal pulse. It was not possible
to check if the variability is coherent. DCL99 conclude ``Although pulsed
emission seems a reasonable assumption, the power excess could arise from
variability of a different nature (e.g. broad-band variability increasing the
chance of spurious detection)."
The orbital period, pulse period and observed X-ray luminosity are remarkably
similar to those of the Small Magellanic Cloud neutron star XRB \object{SMC X$-$1} 
\citep{2000A&AS..147...25L}. 
\x7\ was the first and only identified eclipsing accreting binary system
with an X-ray source in an external galaxy other than the Magellanic Clouds before 
the detection of similar behavior based on \xmm\ and \chandra\ data of the NGC~253 
X-ray source RX~J004717.4-251811 \citep{2003A&A...402..457P}.

\citet[][hereafter PMM2004]{2004A&A...413..879P} analyzed several observations of the 
\xmm\ \m33\ survey and an archival \chandra\ observation where \x7\ was in the
field of view. The observations cover a large part of the 3.45 d orbital period,
however not eclipse in- and egress.  
PMM2004 detected emission of \x7\ during eclipse and a soft X-ray
spectrum of the source out-of eclipse that can best be described by
bremsstrahlung or disk blackbody models. No significant regular
pulsations of the source in the range 0.25--1000~s were found. The
average source luminosity out of eclipse was 5\ergs{37} (0.5--4.5~keV, corrected
for Galactic foreground absorption).
In a special analysis of DIRECT\footnote{For information on the DIRECT project
see {\tt
http://cfa-www.harvard.edu/\~\/kstanek/DIRECT/}.} observations PMM2004
identified as the optical
counterpart a B0I to O7I star of 18.89 mag in V which
shows the ellipsoidal light curve of a high mass X-ray binary (HMXB)
with the \x7\ binary period. Based on the location of the X-ray eclipse and the
optical minima, PMM2004 derived an improved binary ephemeris and argued that the compact
object in the system is a black hole. Those authors reached this conclusion 
based on the mass of
the compact object derived from orbital parameters and the optical 
companion mass, the lack of pulsations, and the X-ray spectrum.
\x7\ would be the first detected eclipsing high mass black hole XRB.

The \chandra\ ACIS-I survey of \m33\ (ChASeM33) is a very large program which
will accumulate in seven deep pointings, each 200 ks in length, 
a total exposure of 1.4~Ms. During several of these pointings
\x7\ is in the field of view. We report here on ChASeM33 observations of 
\x7\ spread over just 20 binary orbits, 
which resolved for the first time the eclipse ingress and egress 
and allowed us to constrain the light curve of \x7\ for binary phases around 
eclipse. Preliminary results of the first
observations were announced by \citet{2005ATel..633....1S}.
In addition we identify the source on archival HST WFPC2
images. Throughout the paper, we assume a distance to \m33\ of 795~kpc 
\citep{1991PASP..103..609V}.

\section{Chandra observations and results}
\x7\ was sampled by the ACIS-S3 chip in one \chandra\ observation
and
by the I0, I1 and I2 chips during 11 additional observations. 
Table~\ref{tbl:obs} summarizes these observations
giving observation identification (ObsID) in column 1, observation start date (2), 
elapsed time
(3), the ACIS chip covering \x7\ (4), the offset of \x7\ from the pointing 
direction (5), and the \x7\ binary phase during the observation (6) using ephemeris that will be
discussed in Sect.~\ref{sec:tim} and \ref{sec:eph}. The source brightness varied from 
$\sim$3\expo{-3} \ct\ to $\sim$0.2 \ct\ normalized to ACIS-I on-axis. 

Of the pointings considered here, 
the Field 5 position of \x7\ is located close enough to the optical axis 
to result in significant pile-up,
particularly during the high phase.  To correct the light curve for 
the effects of pile-up, we used the best-fit model for unpiled data from 
the other pointings (see
Sect.~\ref{sec:spec}), and applied a pileup(phabs(diskbb)) model within
XSPEC.  We estimated the pile-up parameters by fitting simultaneously
the high state data for ObsID 6382 (insignificant pile-up) and 
ObsIDs 7170 and 7171 (significant pile-up).  The phabs and diskbb
models were set to their best fit values from the unpiled spectra;
the diskbb normalizations were tied for all three datasets.
All of the pile-up model parameters except the grade morphing 
parameter $\alpha$ were frozen. For a description of the  pile-up model see 
\citet{2001ApJ...562..575D}.  The
pile-up parameters were set to their default values, but the fr\_time
parameter was frozen at 3.2 for ObsIDs 7070, 7071, and
at 0 for 6382; this has the effect of turning off the pile-up 
model component for ObsID 6382.  The resulting reduced 
$\chi^2$ of the fit was 0.97 and the corresponding value for
$\alpha$ was 0.692$^{+0.114}_{-0.116}$ (90\% confidence limits).  
Using the fitted value for $\alpha$, a correction 
curve was generated by evaluating the ratio of the XSPEC 
``Model predicted rate'' without pile-up (fr\_time=0) to the corresponding 
rate with pile-up (fr\_time=3.2) as a function of the model predicted rate 
(with pile-up).  This ratio was used to correct the observed count rates of 
ObsIDs 6384, 7070, and 7071  for the effects of pile-up.
  
For the spectral analysis we used standard Level 2 event files cleaned for
columns with higher background rates than adjacent columns. 
In the observations of
Fields 4 and 5, counts from \x7\ are rejected when the source is moving across
rejected columns (due to the satellite dithering). This effect creates spurious
periods in period analysis and can reduce the counts in 1000~s integration
intervals by varying amounts (up to more than 30\%). Therefore, we created new
Level 2 event files for broad band time variability analysis that did not
reject these columns. We correct for the detection efficiency at different 
off-axis angles using factors derived from response files for the different CCDs and
off-axis positions assuming the disk-blackbody spectrum derived in 
Sect.~\ref{sec:spec}. 
We normalized the count rates to a CCD 3 on-axis rate. 

The data analysis was performed using tools in the ESO-MIDAS v05SEPpl1.0, 
EXSAS v03OCT\_EXP, CIAO v3.2 and LHEASOFT v5.3
software packages as well as the imaging application DS9 v3.0b6. 

\subsection{Time variability}\label{sec:tim}
During ObsID 6378 and 7171 we observed transitions by \x7\ into and out of
eclipse, respectively. We sampled both background and solar system barycenter corrected 
light curves of \x7\  with a time resolution of 1000~s. To increase the signal to noise
specifically in the far off-axis ObsID 6378, we
restricted the analysis to the 0.5--5~keV band which covers most of the source
flux. Further sub-dividing this energy
band into a hard and soft band did not show any significant hardness ratio 
changes of the in- 
and egress behavior as could be expected based on \exo\ observations of LMC~X$-$4
and Her~X$-$1 \citep[see Figs.~78 and 79 in ][]{1991PhDT.......139D}.
To determine eclipse start and end times, we approximated the light curves
assuming constant count rates within and out-of eclipse and a linear transition
in between using a $\chi^2$ minimization technique. We searched for $1\sigma$ errors 
of the eclipse start and end time, respectively, assuming those to be the only interesting 
parameter for the fit.

In ObsID 6378 the transition into eclipse lasted 12.75~ks and ended at
HJD~245\,3635.4110$\pm$0.0037. The transition out of eclipse in 
ObsID 7171 started at HJD~245\,3642.8272$\pm$0.0052 and lasted
for 10.52~ks. From these ingress and egress times separated by two orbital
periods, we directly derive the mid eclipse ephemeris of the eclipse in between
to HJD~$245\,3639.119\pm0.005$. Assuming this epoch as phase zero and a
binary period of 3.453014~d (see Sect.~\ref{sec:eph}), we calculated 
light curves of all \x7\ observations (Fig.~\ref{fig:lc}). 
We also determine an eclipse duration of less than
0.147$\pm$0.006 in phase corresponding to an eclipse half angle of 
$26.5\degr\pm1.1\degr$.

ObsIDs 7196 and 7199 fully fall into eclipse. ObsID 6384 at phase
0.83 to 0.90 indicates a much longer transition into eclipse than ObsID 6378.
ObsID 7198 shows dipping behavior well before eclipse with a return to
the out-of eclipse intensity level at the end. ObsID 6382 shows a second
egress from eclipse which is significantly faster than the one
16 orbits earlier (ObsID 7171). However, the observation starts during egress and
the phase range within eclipse is not covered. ObsID 7226 covers eclipse
ingress at the end of the same orbit and again shows strong dipping well before
the ingress. Eclipse egress and ingress times are consistent with the
times derived above. Generally speaking the variability of \x7\ before
eclipse (phase 0.7 to 0.9) seems to be much more pronounced in individual observations 
than after eclipse (phase 1.1 to
1.4). Average count rates in eclipse and out of eclipse are 0.003~\ct\ and 
0.15~\ct, respectively. This out-of-eclipse count rate varies in different 
binary orbits by factors of 1.3 and there are residual short term fluctuations
that can be described by dips with a similar amplitude and a duration of
several 1000~s also outside the pre-eclipse phase.
 
To search for pulsations we extracted light curves in the 0.5--5~keV band from 
the longest observations after eclipse, ObsIDs 6382 and 7170. We created power density 
spectra in the frequency range of 10$^{-4}$--0.15~Hz and found no significant 
periodic signal with a 3$\sigma$ upper limit of 5.3\% for sinusoidal variations. A power 
density spectrum derived from ObsID 6382 by adding the power spectra from 
23 intervals of 3319 s length (1024 time bins with the instrument resolution of 
3.241 s) is shown in Fig.~\ref{fig:pds}.
The power spectrum is flat at a value of 2.

\subsection{Energy spectra}\label{sec:spec}

We analyzed energy spectra of \x7\ for all observations. For ObsIDs 6378 
and 7171, times out of eclipse, during ingress and egress, and in eclipse were 
handled
separately. Absorbed power-law, bremsstrahlung and disk-blackbody models 
(which were found to best represent the \xmm\ and \chandra\ spectra 
analyzed by PMM2004) were first fit to the individual spectra. The power-law fit to 
the spectrum obtained from ObsID 6382 (which has the highest statistical quality)
yields an unacceptable fit with a reduced $\chi^2$ of 1.92 while bremsstrahlung and 
disk-blackbody models result in $\chi^2_r$ of 1.45 and 1.32, respectively.
The derived parameters are consistent within the errors for all spectra including the 
spectra accumulated during eclipse ingress and egress and during ObsID 7198 
when the source shows high variability. Therefore, we performed a simultaneous fit
with the disk-blackbody model to the spectra of eight out-of-eclipse observations 
excluding times when the X-ray source was in eclipse and forcing the absorbing 
column density to be the same for all observations (this reduces the number of free
fit parameters compared to the individual fits). The resulting inner disk 
temperature was systematically higher (at $\sim$1.3 keV) for ObsIDs 6384, 
7170 and 7171: these are the observations 
which include \x7\ nearly on-axis. The most likely 
reason for the ``harder'' spectra is pile-up and we therefore excluded these three 
spectra from further analysis. The fit to the remaining five spectra from ObsIDs 
6378, 6382, 6386, 7197 and 7198 shows no significant differences in the inner disk 
temperature and therefore this parameter was also forced to be the same in the 
simultaneous fit. We refit the power-law and bremsstrahlung models
for comparison. The best fit is obtained with the disk-blackbody model yielding
a $\chi^2_r$  of 1.10. The $\chi^2_r$  for the best fitting bremsstrahlung and
power-law models are 1.16 and 1.44, respectively.
The derived spectral parameters are similar to the ones reported by PMM2004:
disk-blackbody with inner disk temperature kT = 0.99$\pm$0.03~keV and 
N$_H$ = (0.95$\pm$0.10)$\times 10^{21}$ cm$^{-2}$;
bremsstrahlung with temperature kT = 2.74$\pm$0.13~keV and
N$_H$ = (2.05$\pm$0.12)$\times 10^{21}$ cm$^{-2}$;
power-law with photon index $\gamma$ = 2.38$\pm$0.05 and
N$_H$ = (3.32$\pm$0.17)$\times 10^{21}$ cm$^{-2}$.
The best fit disk-blackbody model is shown in Fig.~\ref{fig:spec}.

The normalization of the disk-blackbody model is given by 
K = $(r_{\rm in}/{\rm d})^2$(cos $i$) with the inner disk radius 
r$_{\rm in}$ in km,
the source distance d in units of 10 kpc and the disk inclination $i$. 
The spectra show variations by a factor of $\sim$2 in normalization with
K=0.054 for ObsID 6378 and relative factors 0.58, 0.94, 1.13 and 1.37 for 
ObsIDs 7198, 7197, 6386 and 6382, respectively. 

The N$_H$ values for the models discussed above clearly indicate absorption 
within \m33\ or intrinsic to the source in addition to
the Galactic value \citep[5.86 and 6.37 $\times 10^{20}$ cm$^{-2}$ in the direction
of \x7\ according to][respectively]{1990ARA&A..28..215D,1992ApJS...79...77S}.
Absorbed and unabsorbed source fluxes in the 0.3--10 keV band are in the range 
(5.4--12.6)\ergcm{-13} and (6.2--14.7)\ergcm{-13}, respectively,
based on the best fitting disk-blackbody model. These fluxes correspond to
source luminosities of (4.1--9.6)\ergs{37} and (4.7--11.2)\ergs{37}, 
respectively. 

The  N$_H$ value of the best fitting disk-blackbody model indicates that
\x7\ lies on the near side of \m33 as the absorbing column within \m33\ can 
be determined to $\sim$2.2\hcm{21} from a
$47\times 93$ arcsec half power beam width H{\sc i} map 
\citep{1980MNRAS.190..689N}. From the N$_H$ value we can 
compute the expected optical extinction $A_{\rm V} = 0.53\pm0.06$ mag 
and $E(B-V) = 0.18\pm0.02$ using the standard relations
\citep{1995A&A...293..889P}. These numbers are in the range given by 
PMM2004 who assumed that we see \x7\ through less than half the absorbing 
column within \m33.

\subsection{Improved position}
\x7\ was located in four of the five ChASeM33 fields observed so far.
The source was closest to on-axis (1.58\arcmin\ off-axis) in
the Field 5 observations (see Table~\ref{tbl:obs}).  
This results in the most compact PSF
and thus the most reliable position determination. To
refine the absolute astrometry of the Field 5 data, we searched
the USNO-B1.0 and 2MASS catalogs for close positional matches with
X-ray sources. Similarly, to improve the statistics for X-ray centroiding,
we worked with a merged dataset for ObsIDs 6384, 7170, and 7171.
We identified 9 candidate optical/2MASS objects.  Six were rejected 
because of far off-axis positions or small number of counts.  The 
remaining candidates were $\ge 7^\prime$ off-axis, 
for a 2MASS object ($\sim$250 counts, 0.5--5~keV) which was
7.8\arcmin\ off-axis. We enhanced the number of candidates for registration by
adding two isolated centrally brightened supernova remnants (SNRs) with good radio positions 
\citep[sources 57 and 64 from the list of][]{1999ApJS..120..247G} assuming that
the finite size of SNRs -- and
potential differences in the X-ray versus radio distribution -- 
do not bias the position determination.
We determined the X-ray centroids based on an 
iterative sigma-clipping algorithm applied to the 0.5--5~keV X-ray data.
Based on an initial position estimate and clipping radius, the
standard deviation of the radial distribution is evaluated, and
points greater than a given number of standard deviations are rejected.
The iteration of centroiding and rejecting events continues until the 
centroid converges to within a specified tolerance or for a fixed 
number of iterations (10).  The difference in sky coordinates between
the catalog position and the X-ray centroid position was evaluated
for each source.  The mean offset was $\Delta x=(0.23\pm0.37)\arcsec, 
\Delta y=(0.32\pm0.21)\arcsec$.  The centroid of the \x7\ source was evaluated
in the same way, and the resulting offset was applied to correct
the sky position.  Finally, the corresponding celestial coordinates
were evaluated using the CIAO tool dmcoords; the aspect solution
files (asol1 files) were used in order to correct the positions
using the aspect offsets.  The resulting \x7\ position is
RA$_\mathrm{J2000} = 01^h33^m34\fs12, \delta_\mathrm{J2000} =
+30\degr32\arcmin11\farcs6$
with a combined error of 0.5\arcsec. The position is within 2$\sigma$ of that
given by PMM2004 based on a registration using just one SNR.

\section{Optical observations and results}
\mx7\ is located along the line of sight to the
dense OB association HS 13 \citep{humphreys1980_m33_h2reg}: it was 
identified by PMM2004 with a
specific star within this association, as a result of detection of
regular (ellipsoidal) variations in B and V at the X-ray period. The
OB association has been imaged with {\it HST} using WFPC2 in three
filters:  F336W, F439W, and F555W.  The observing details are listed
in Table \ref{tab_hst}.  In an attempt to learn more about the
optical counterpart to \x7, we retrieved relevant datasets from the
MAST archive.  Data retrieved from the archive are automatically
reprocessed with the latest calibration files.  After inspecting the
individual exposures, we combined the two exposures taken with the
F336W filter and the two taken in F439W (using the STSDAS task
gcombine) to eliminate the effects of cosmic rays on the images.

The position of the X-ray source as determined with \chandra\ is shown
in Fig.~\ref{fig:opt}. Since the error in the astrometric solution of images
in the HST pipeline is typically $\sim$1.5\arcsec\,--2\arcsec, we 
attempted to reduce the positional uncertainty by registering the WFPC2 
images to the USNO-B1.0 or 2MASS frames.  Given the small field of view and
greatly superior spatial resolution of the WFPC2 images, many 
sources identified in the USNO-B1.0 or 2MASS catalogues are resolved into 
multiple objects, making a unique match between stars in the image and
sources in the catalogue difficult.  Therefore, we performed our 
astrometric corrections in two steps, taking advantage of the 
much larger field of view of the KPNO Mosaic B image of \m33\ 
\citep{massey2002}.  First, we identified a sample of 25 isolated, bright USNO-B1.0
stars within 4\farcm5 of \x7\ to compute the shift required to bring
the Mosaic B image into the USNO-B1.0 frame.  The RMS positional error
of these stars in the corrected Mosaic image was unacceptably large (0\farcs85).
We then tried the same procedure using 21 bright 2MASS stars located in isolated
areas in the  KPNO Mosaic B image. The RMS positional error
of these stars in the corrected Mosaic image was 0\farcs16.
We then used 10 bright,
isolated stars in common between the Mosaic B and F439W images (restraining the
selection to the region covered by the WFPC2 CCD in which the \x7\ counterpart 
was located) to compute the shift
required.
The RMS positional error in the shifted HST image was 0\farcs13.  Combining
the 0\farcs1 absolute uncertainty of the 2MASS position with the
uncertainties listed above, the absolute astrometric uncertainty of the
final registered F439W image is 0\farcs23.

We applied the calculated shifts above to the HST image. 
Fig.~\ref{fig:opt}  shows a 10\arcsec\ $\times$ 10\arcsec\ field from the F439W image
centered on \mx7. The black solid circle shows our best estimated position
for the X-ray source from the \chandra\ image.  The error circle of \x7\
is 0\farcs5 in radius (see above). The HST positional accuracy is indicated by a
cross. We also show the error quoted by PMM2004 in red. The error circles
overlap with a bright star, coincident with that proposed by PMM2004 (based also
on time variation arguments) as the donor star for the compact object. 

We also carried out aperture photometry of the stars in the
{\it HST} field (using the IRAF procedures daofind and phot).
We find that the optical counterpart to \x7\ 
with the WFPC2 resolution is not a blend of stars. 
PSF fits to the source in the F336W, F439W, and F555W images give FWHM 
compatible with the other point-like sources in the images. The star 
about 0\farcs9 to the South on the other hand is just resolved in at least 
two equally
bright sources which are separated by $\sim$0\farcs2. We can rule out the
presence of another star of similar brightness at the position of the optical
counterpart of \x7 to this distance, which corresponds to a projected separation
at the distance of \m33\ of 0.8~pc. 

The optical counterpart has 
apparent magnitudes of 17.6, 18.2, and 18.9 for the F336W, F439W, and
F555W filters, respectively, in the STMAG system.  The colors derived from these filters
for the counterpart to \x7\  are typical of the other bright
stars ($m_{\rm V}<20.5$) in HS 13 as observed with WFPC2. The F336W, F439W,
and F555W filters are centered approximately on the corresponding U, B
and V filters. Adopting a color
transformation of 0.5~mag for U, 0.66~mag for B, and 0.03~mag for V
\citep{2002datahandbook}, we find $m_{\rm U}$ of 18.1~mag, 
$m_{\rm B}$ of 18.8~mag, and $m_{\rm V}$ of 
18.9~mag (i.e. U--B of -0.7~mag, B--V of -0.1~mag) for the optical 
counterpart to \x7. 

\section{Discussion}

The well sampled orbital light curve of \x7\ indicates stronger variability 
before eclipse compared to after eclipse 
(Fig.~\ref{fig:lc}). Variability at this phase is often observed in HMXBs and 
is explained by the viewing geometry through the innermost regions of the
wind of the companion and dense material following the compact object in its orbit 
\citep[e.g.][]{1992A&A...263..241H}. Dense structures are created by the gravitational 
and radiative interactions of the compact object with the stellar wind
\citep{1990ApJ...356..591B,1991ApJ...371..684B}. 
This behavior is also reflected in the on-average
longer eclipse ingress time and longer eclipse duration derived by DCL99. 

Similarly to
PMM2004 we find residual emission from the source during eclipse. 
Residual emission during eclipse was measured from most 
eclipsing XRBs and can be explained by re-processing
of primary photons from the compact X-ray source in an extended accretion
disk corona (which is not fully occulted) or by scattering in the companion 
atmosphere/stellar wind. Residual emission of up to $\sim$10\% of the
uneclipsed flux was reported 
\citep{1991A&A...252..272H,1992ApJ...389..665L,1996PASJ...48..425E} 
depending on system geometry and wind density. The
\x7\ residual emission of  $\sim$4\% is well within these limits.

\subsection{Improved ephemeris}\label{sec:eph}
DCL99 modeled the folded light curve from \x7\ as a constant flux plus linear ingress and
egress plus an eclipse interval with zero flux. It is obvious from their data (see
their Fig.~1) that the eclipse egress is better determined than the eclipse center and
duration. These strongly depend on the shape of the pre-eclipse dips contained in the
light curve as can be seen from the resolved pre-eclipse behavior in the \chandra\ data
(Fig.~\ref{fig:lc}). 
DCL99 did not determine the eclipse parameters from individual
eclipses but only from the average light curve due to limited statistics. 
The time of eclipse egress is the best determined parameter.
Unfortunately, in the paper they do not give this parameter separately but only the
center of eclipse and length of eclipse. In the following we use eclipse egress 
times to determine an improved orbital period $P$ and a possible period 
derivative $\dot{P}$.

Due to limited phase coverage, the \xmm\ observations do not resolve  eclipse ingress or
egress. The time of eclipse egress can only be constrained to $<$0.02 in phase. 
Based on these data PMM2004 restricted the time of eclipse egress to
HJD~$245\,1760.953\pm0.035$ (note typographical error in egress time in PMM2004, 
but the calculation
used the number given here) and determined a time of mid-eclipse assuming the 
eclipse shape parameters of DCL99. With this mid-eclipse epoch and the one given 
by DCL99, PMM2004 determined an improved orbital period.

The maximum eclipse duration of 0.147$\pm$0.006 (Sect.~\ref{sec:tim}) determined from
individual observations is -- as expected -- significantly
shorter than the one given by DCL99 (0.20$\pm$0.03) from average orbit fitting. 
From the DCL99 ephemeris it is not possible to derive a well defined eclipse 
egress time. The same is true for the \ein\ results 
\citep[][hereafter PRC89]{1989ApJ...336..140P}. 
We therefore decided to re-analyze relevant \ro\ and \ein\ data.

\ro\ PSPC observations of \m33\ were short compared to the HRI observations and 
did not cover the \x7\ eclipse egress. We therefore restricted our re-analysis
to
\ro\ HRI data. After screening for high background, we combined data with
continuous observation intervals which lead to variable integration times of 
typically 1800~s (minimum 41~s, maximum 3791~s) depending on the duration of the
scheduled observation and background. DCL99 in contrast grouped the 
data in 3000~s averages. Only once is the eclipse egress closely monitored.
In the \ro\ interval corresponding to
ObsID 600488h, X-7 was still in eclipse while in the following interval
(corresponding to ObsID 600489h), less than 0.04d later, X-7
already featured a count rate that indicated the source was out
of eclipse. This allowed us to restrict the time of eclipse egress to
HJD~$244\,9571.724\pm0.018$. 
In another case during ObsID 600020h-1, observations in eclipse and out of
eclipse are separated by 0.26 d, which does not allow us to further constrain
eclipse egress times.    

PRC89 reported \ein\ IPC and HRI observations of \x7.
For the HRI observations PRC89 combined several continuous observation
intervals to get significant data. For the IPC observations PRC89 simply
integrated over individual continuous observations. Due to the better statistics
in the IPC observations we considered only IPC eclipse egress coverages 
(i.e. ObsID I2090, see also PRC89, Fig.~1). In images of the observation
\x7\ is only visible in the energy band 0.6--2.8~keV (PI 4--9).
We therefore restricted the analysis to this energy band. We selected
extraction position and area by comparison with the close-by bright central source 
X$-$8. We specifically investigated the
``first rising episode at day 1.5" as identified by PRC89. In contrast to the
report by PRC89, \x7\ count rates
during this period do not show increasing flux
but the source intensity is compatible with zero during all three intervals.
In the next set of observation intervals at
around day 2 the source is clearly out of eclipse. This indicates an eclipse 
egress between  HJD~244\,4087.840 and HJD~244\,4088.255. 

Combining the \ro\ eclipse egress boundaries with the \chandra\ ephemeris
suggests an 
orbital period of 3.453014$\pm$0.000020~d. The lower panel of Fig.~\ref{fig:lc}
shows the \ro\ HRI light curve folded with the above period and assuming the
\chandra\ mid-eclipse epoch, for the same phase range as the \chandra\ data 
above. As can be seen in Fig.~\ref{fig:orb}, these ephemerides also are 
consistent with the boundaries determined for the \xmm\ eclipse egress. However, 
they seem to miss the boundaries determined above for the \ein\ IPC observation. 
If we assume a constant rate of change of the orbital period over this time, we 
can model all eclipse egresses. The parabola shown in Fig.~\ref{fig:orb}
assumes a period of 3.45294~d at the \chandra\ epoch and orbital period decay 
rate of $\dot{P}_{\rm orb}/P_{\rm orb} = -4\times10^{-6}$ yr$^{-1}$. It nicely
models the average eclipse egress times of \xmm, \ro\ and \ein. However, also
a small $\dot{P}_{\rm orb}/P_{\rm orb} = -0.7\times10^{-6}$ yr$^{-1}$ and a period of 
3.45302~d or a much higher  
$\dot{P}_{\rm orb}/P_{\rm orb} = -7.5\times10^{-6}$ yr$^{-1}$ and a period of 
3.45285~d would still be consistent with all the eclipse egress boundaries.

The derived orbital decay for \x7\ is well within the range of values 
determined for other HMXBs like 
Cen X$-$3
\citep[$(-1.738\pm0.004)\times10^{-6}$ yr$^{-1}$,][]{1992ApJ...396..147N},
SMC X$-$1
\citep[$(-3.36\pm0.02)\times10^{-6}$ yr$^{-1}$,][]{1993ApJ...410..328L}
or LMC X$-$4 
\citep[$(-9.8\pm0.7)\times10^{-7}$ yr$^{-1}$, see e.g.][]{1996ApJ...456L..37S,2000ApJ...541..194L}. 
Such rapidly decreasing periods in HMXBs are most likely caused by tidal
interaction between the compact object and its massive companion. As the orbit
decays the Roche lobe will descend into the companion's atmosphere and mass
transfer will increase to super-Eddington rates over a relatively short
time scale. In the end, the compact object is expected to 
spiral into the envelope of the companion and in this way terminate the  high-mass XRB
phase of the evolution \citep[see e.g.][]{1993ApJ...410..328L}.

\subsection{The optical companion}
The HST WFPC2 images clearly resolve the 
dense OB association HS 13 \citep{humphreys1980_m33_h2reg}. The optical 
counterpart is located to the North and is one of a pair of stars with similar
luminosities: these stars are separated by $\sim$0."9. While the
suggested counterpart is presumed to be a single source, the source to the South
is a blend of at least 2 stars (elongation from SSE to NNW).
The HST observations were carried out within one hour on October 25, 1995
corresponding to binary phase 0.76 using the ephemeris given in
Sect.~\ref{sec:eph}. Phase 0.76 corresponds to the second maximum of the 
ellipsoidal light curve. Assuming the parameters of the optical light curve by
PMM2004 (sinusoidal fit with 0.033 mag amplitude) the correction to X-ray
eclipse is +0.066 mag resulting in  corrected magnitudes $m_{\rm B}$ of 18.9 mag, 
and $m_{\rm V}$ of 19.0 mag for the optical counterpart to \x7.  
Magnitudes and colours are very close (within 0.1 mag)
to values deduced by PMM2004. \x7\ is also included in the UBVRI photometry of
stars in \me31\ and \m33\ from the survey of Local Group galaxies, recently
published by \citet{2006astro.ph..2128M}. They give a position (registered with
the USNO B1.0 frame) of RA$_\mathrm{J2000} = 01^h33^m34\fs18, 
\delta_\mathrm{J2000} = +30\degr32\arcmin11\farcs5$, about 0\farcs3 ESE from the
position in our registered HST image position (see Fig.~\ref{fig:opt}). Their
\x7\ magnitude and colours ($m_{\rm V} = 19.087\pm0.007$mag, $m_{\rm (B-V)} =
-0.084\pm0.10$mag, $m_{\rm (U-B)} =-1.057\pm0.11$mag) coincide with those
determined by us within 0.2 mag. Part of the discrepancy may be due to the
sampling at different binary phases.

The type and luminosity class of the star can be deduced from the absolute
optical magnitude and colour during eclipse when we see the optical surface that
is mostly undisturbed by gravitational effects, an expected accretion disk and
heating by the X-ray source. To derive the absolute magnitude the measured
brightness has to be corrected for the distance (-25.50~mag for the assumed
distance of 795 kpc) and for interstellar extinction, the colour has to be
corrected for reddening. These corrections have been estimated from the
absorption of the X-ray spectrum in Sect. 2.2. Based on the HST values the  
companion star should have an absolute  $M_{\rm V}$ of $-6.1$ mag and $(B-V)_0$ of
$-0.3$ mag. This corresponds to a giant star of spectral type O6III which
has a temperature of 39\,500~K, 
a radius of 17 $R_{\sun}$ and a mass well above 20 $M_{\sun}$ 
\citep[see Appendix E in][]{1996QB461.C35......}.

\subsection{M~33 X$-$7, an eclipsing black hole HMXB}

With the new eclipse duration and the better determination
of the companion type on the basis of the extinction combined with the colour 
excess corrections derived from 
the absorbing column of the X-ray spectrum of \x7, we can significantly 
improve the mass estimate of the compact object compared to PMM2004. To do so we
correct the B and V light curves given in PMM2004 for extinction and colour 
excess. We then model these light curves using the ``PHysics Of Eclipsing BinariEs"
program PHOEBE \citep{2005ApJ...628..426P} built on top of the widely used WD
program 
\citep{1971ApJ...166..605W,1979ApJ...234.1054W,1990ApJ...356..613W}.
By adjusting the semi-major axis of the binary system we kept the radius of the
secondary star at 17$R_{\sun}$. We also fixed the temperature of the
secondary to 39\,500~K. We then fitted the luminosity of the secondary and the 
mass ratio of the companions using the light curves in the B and V band
simultaneously, assuming different inclination angles. We derive acceptable fits
for inclination angles from 75\degr\ to 90\degr. Angles of 74\degr\ and
smaller can be excluded as the companion 
star would overfill its Roche lobe. The derived parameters are given in
Table~\ref{tab_fit}.  The available optical data are not sufficient to 
discriminate between the allowed inclination angles. However, we expect that 
an inclination above 80\degr\ is more likely than an inclination in the range 
75\degr\ -- 80\degr\ where the companion nearly fills its Roche lobe
and unstable mass transfer would be expected based on the X-ray observations. 
In the case of unstable mass transfer, the X-ray emission from \x7\ would most likely be much 
more variable on longer time scales than observed in all observations since 
those with the \ein\ observatory. This implies a mass of
the compact object in the system  of greater than 9$ M_{\sun}$ and
clearly indicates a black hole as the compact object in the system.

As already discussed in PMM2004, further arguments for the black hole nature of the 
compact object in \x7\ come from the lack of X-ray pulsations, the short term
variability and the X-ray spectra. We discuss each of these
properties in turn:

(I) Pulsations are clear indicators of a neutron star as the compact object in a
HMXB. In the power density spectrum analysis we did not detect significant
periodic signals allowing a black hole as the compact object. However, this does
not rule out the possibility that the compact object is a neutron
star. Our unsuccessful \chandra\ periodicity search was limited towards
short periods by the ACIS-I sampling time of 3.2~s. Also, with
the significantly shorter sampling time of \xmm\ EPIC pn of 0.073~s, PMM2004 did
not find significant pulsations.

(II) Low accretion rate
sources should show a power density spectrum (PDS) with a broken power law
\citep[see][and references therein]{2005astro.ph..8284B}. Above a luminosity of
typically 0.1 of the Eddington luminosity, the PDS should be flat.
The short term fluctuations seen in the power spectral analysis of \x7\ are 
very small as one would expect for such a high accretion rate source. The
unabsorbed luminosity of \x7\ in the 0.3--10~keV band of $>$1.1\ergs{38} 
in maximum is consistent 
with a stellar mass black hole, but it does not constrain the mass.

(III) HMXBs with a neutron star as the compact object normally show power law spectra 
(photon index 0.8--1.5) with a high-energy cutoff around 10--20 keV 
\citep[see e.g.][]{1983ApJ...270..711W,2000ApJ...535..632M}. 
Disk-blackbody spectra, on the other
hand, suggest emission that is dominated by the inner accretion disk of a low
mass XRB
system, or -- in case of a HMXB -- the presence of an accretion disk surrounding
a black hole emitting in the high state \citep[e.g.][]{1986ApJ...308..635M}. 
As mentioned before, the variability of \x7\ outside eclipse most likely is 
caused by partial covering of the X-ray emission region 
due to material in the accretion stream or in the outer accretion disk.
If we assume that during the brightest phase, the \x7\  disk-blackbody
normalization corresponds to the innermost stable circular orbit with
radius $r$ around a black hole, i.e. 
\hbox{$r = 3 R_s = \EXPN{9}{5} (M_x/M_{\sun})$ cm} (with $R_s = 2GM_x/c^2$ the
Schwarzschild radius), we obtain $M_x > 2.4 M_{\sun}$. 

All of these results of the new \chandra\ observations suggest that \mx7\ 
is the first known eclipsing HMXB with a black hole as the compact object.

\citet{2005ApJ...634L..85P} advanced the idea of searching for
eclipses in ultra-luminous
X-ray sources (ULXs) to determine black hole masses and -- most importantly -- 
to separate
intermediate mass from stellar mass black holes as the compact object in this
exceptional class of X-ray sources. 
They proposed to compare the number of eclipses
in the different kind of systems and predicted that more eclipses by far
should be detected in stellar mass systems
than in intermediate mass black hole
systems.The orbital periods and other system parameters would provide
considerable insight as to the nature of the binary. While M33 X-7 is about a factor of five
less luminous than the commonly accepted lower limit of ULX
luminosities, it is still the first eclipsing black hole
XRB system and -- if in other observations detected in a factor of five 
brighter phase -- may be the first stellar mass ULX candidate suitable
for the proposed type of investigations.

As we discussed above, the resolved eclipse ingress and egress in the \chandra\ 
energy band is not caused by the extent and the structure of the X-ray emitting
region but by absorption in the density structure of the atmosphere of 
companion and the accretion stream. \citet{2005PASJ...57..827W} investigated the 
asymmetry of black-hole accretion flows and suggested the presence of 
effects with time scales of seconds due
to relativistic effects close to the black hole. These should lead to asymmetries in
the eclipsing light curves on time scales of seconds. The \chandra\ observations
of the eclipse in- and egress of \x7\ show that such effects are masked by much
longer time scale effects discussed above. They may only be observable by
sensitive observations at higher X-ray energies such as those where absorption 
effects in the companion atmosphere and the accretion flow vanish.  

\section{Conclusions}
The ChASeM33 observations clearly resolved eclipse ingress and egress of the 
persistent eclipsing HMXB \mx7. Combining the eclipse egress times with 
\ein, \ro, and \xmm\ observations allowed us to improve on the orbital period 
which indicates 
orbital decay. The X-ray spectrum can best be described by a disk-blackbody
spectrum with parameters that do not change significantly in different
observations while the source luminosity changes by up to a factor of two.
The short term variability of \x7\ can be described by a flat power density 
spectrum. No significant regular pulsations 
were found in the frequency range of 10$^{-4}$--0.15~Hz.
HST WFPC2 images resolve the optical counterpart, which with the help of extinction
and colour corrections derived from the X-ray absorption can be identified as an
O6III star. The X-ray period and eclipse
duration -- together with fits to the light curve by PMM2004 of the
optical companion -- 
imply a black hole with a mass above $9 M_{\sun}$ as the most likely compact 
object in the system. \mx7\ is the first eclipsing black hole HMXB.

\acknowledgments
This research has made use of data obtained through the High Energy Astrophysics
Science Archive Research Center Online Service, provided by the NASA/Goddard 
Space Flight Center, of the USNOFS Image and Catalogue Archive
operated by the United States Naval Observatory, Flagstaff Station
(http://www.nofs.navy.mil/data/fchpix/), 
of data products from the Two Micron All Sky Survey, 
which is a joint project of the University of Massachusetts and the Infrared 
Processing and Analysis Center/California Institute of Technology, funded by 
the National Aeronautics and Space Administration and the National Science 
Foundation. The authors thank Michael Bauer for his introduction to the
``PHysics Of Eclipsing BinariEs" program PHOEBE. 
We thank the anonymous referee for his/her comments and suggestions for
improving the manuscript. 

{\it Facilities:} \facility{CXO (ACIS)}, \facility{HST (WFPC2)}, 
\facility{ROSAT (HRI)}, \facility{EINSTEIN (IPC)}

\begin{table}
\caption{\chandra\ observations of the ChASeM33 program covering M~33~X$-$7.
\label{tbl:obs}}
\begin{tabular}{rrrrrrrr}
\tableline\tableline
ChASeM33 & ObsID & Obs.dates\tablenotemark{a} & Elapsed time & ACIS & Offax &  \multicolumn{2}{c}{X$-$7 binary} \\
Field & &  & (ks) & CCD-ID & (') & phase\tablenotemark{a} & cycle\tablenotemark{b} \\
\tableline
2 & 6378 & 2005-09-21 & 112 & 7 & 17.9 & 0.675--1.044 & -2,-1 \\
5 & 7170 & 2005-09-26 &  41 & 1 &  1.6 & 0.200--0.334 &   0 \\
5 & 7171 & 2005-09-29 &  38 & 1 &  1.6 & 0.041--0.165 &   1 \\
5 & 6384 & 2005-10-01 &  22 & 1 &  1.6 & 0.832--0.905 &   1 \\
6 & 6386 & 2005-10-31 &  15 & 0 &  7.2 & 0.365--0.412 &  10 \\
6 & 7196 & 2005-11-02 &  23 & 0 &  7.2 & 0.928--1.002 & 10,11\\
6 & 7197 & 2005-11-03 &  13 & 0 &  7.2 & 0.144--0.185 &  11 \\
6 & 7198 & 2005-11-05 &  22 & 0 &  7.2 & 0.798--0.868 &  11 \\
6 & 7199 & 2005-11-06 &  15 & 0 &  7.2 & 0.017--0.063 &  12 \\
6 & 7208 & 2005-11-21 &  12 & 1 &  7.3 & 0.622--0.660 &  16 \\
4 & 6382 & 2005-11-23 &  73 & 2 &  9.7 & 0.071--0.312 &  17 \\
4 & 7226 & 2005-11-26 &  25 & 2 &  9.7 & 0.841--0.925 &  17 \\
\tableline
\end{tabular}
\tablenotetext{a}{date of start of observation}
\tablenotetext{b}{with respect to eclipse center HJD~245\,3639.119 and 
orbital period 3.453014~d
(see text)}
\end{table}
\begin{center}
\begin{table}
\caption{HST Observation Log of the M~33~X$-$7 field.\label{tab_hst}} 
\begin{tabular}{ccccc}
\tableline\tableline
{Dataset} & {Date} & {Time} & {Exposure} & {Filter} \\
{~} & {~} & {~} & {(s)} & {~}\\
\tableline
u2tr0301t &  1995-10-25 &  13:40:16 &  160 &  F555W \\
u2tr0302t &  1995-10-25 &  13:45:17 &  400 &  F336W \\
u2tr0303t &  1995-10-25 &  13:54:17 &  400 &  F336W \\
u2tr0304t &  1995-10-25 &  14:04:17 &  500 &  F439W \\
u2tr0305t &  1995-10-25 &  14:15:17 &  300 &  F439W \\
\end{tabular}
\end{table}
\end{center}
\begin{center}
\begin{table}
\caption{M~33~X$-$7 binary parameters for different inclinations determined with
PHOEBE.\label{tab_fit}} 
\begin{tabular}{ccccc}
\tableline\tableline
{Inclination} & {SMA\tablenotemark{a}} & {q\tablenotemark{b}} & {M$_{\rm opt}$} & {M$_{\rm x}$} \\
{(degree)} & {($R_{\sun}$)} & {($M_{\sun}$)} & {($M_{\sun}$)}\\
\tableline
90 &  38.1 &  3.68 &  49.1 &  13.3 \\
85 &  37.4 &  3.88 &  47.0 &  12.1 \\
80 &  35.8 &  4.53 &  42.4 &   9.4 \\
75 &  33.5 &  5.78 &  36.2 &   6.2 \\
\end{tabular}
\tablenotetext{a}{semi-major axis of binary system}
\tablenotetext{b}{mass ratio (Secondary over Primary)}
\end{table}
\end{center}

\clearpage
\begin{figure*}
\includegraphics[height=15cm,bb=98 26 440 700,angle=-90,clip]{f1a.eps}
\includegraphics[height=15cm,bb=140 26 460 700,angle=-90,clip]{f1b.eps}
\caption{Light curve of the X-ray binary M~33~X$-$7.
(Above:) \chandra\ ACIS light curve in the 0.5--5.0~keV band. Individual
observations are marked by their ObsID.
(Below:) \ro\ HRI light curve (0.1-2.4~keV). Observations around eclipse
on JD~244\,9571 are marked as red squares. \label{fig:lc}}
\end{figure*}

\begin{figure}
\includegraphics[height=8cm,angle=-90,clip]{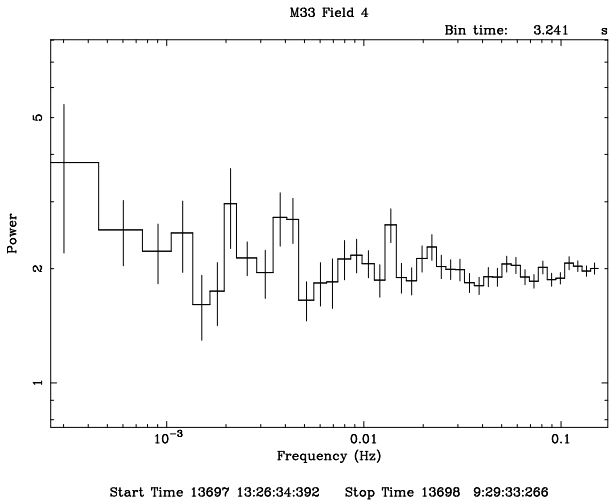}
\caption{\chandra\ ACIS power density spectrum of M~33 X$-$7 from ObsID 
6382 using 23 intervals
of 1024 3.241~s bins.\label{fig:pds}}
\end{figure}

\begin{figure}
\includegraphics[height=8cm,angle=-90,clip]{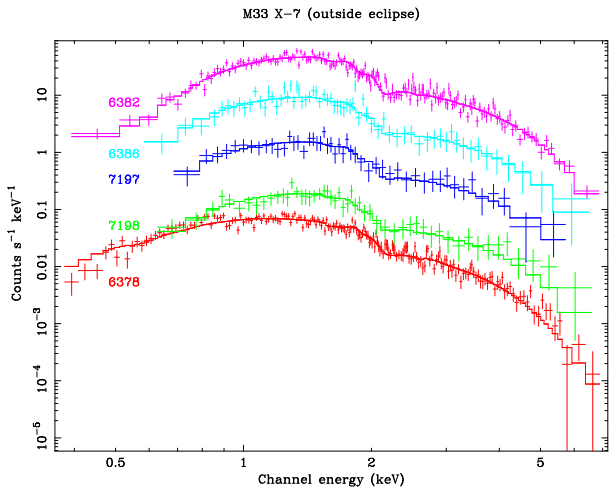}
\caption{\chandra\ ACIS spectra of M~33 X$-$7 during high state. Data and the 
corresponding disk black body model are shown (see text). For a better
representation the spectra were multiplied successively by factors of 1, 5, 5$^2$, 5$^3$ and
5$^4$ from the bottom to the top.
\label{fig:spec}}
\end{figure}

\begin{figure}
\includegraphics[height=8cm,angle=-90,clip]{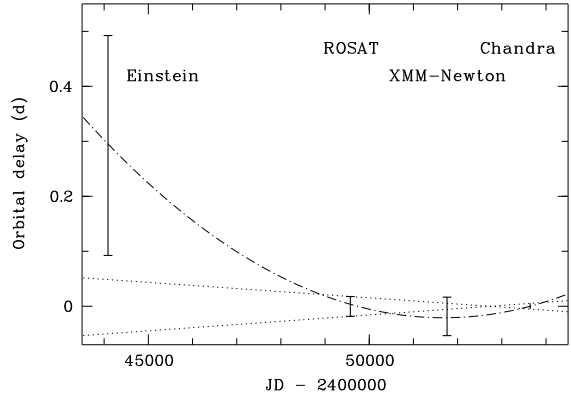}
\caption{Deviation of eclipse egress times relative to best orbital period of  
3.453014~d and \chandra\ eclipse egress ephemeris.
The dotted lines give the shortest and longest periods allowed by the \chandra\ 
and \ro\ data.
The parabola 
assumes a period of 3.45294~d at the \chandra\ epoch and orbital period decay 
rate of $\dot{P}_{\rm orb}/P_{\rm orb} = -4\times10^{-6}$ yr$^{-1}$
(see text).\label{fig:orb}}
\end{figure}

\begin{figure}
\includegraphics[height=8cm,bb=130 225 445 535,clip]{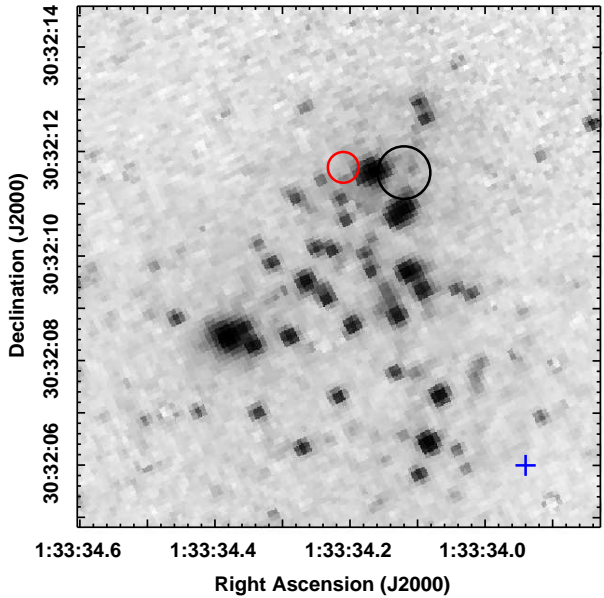}
\caption{HST WFPC2  F439W 10\arcsec$\times$10\arcsec\ image of the HS 13 field
with X-ray positions overlaid. The black circle gives
\chandra\ position and error from Field 5. The red circle indicates the position given by
PMH2004 based on just one radio SNR. Image registration uncertainty is indicated
by the cross in the lower left corner.  
\label{fig:opt}}
\end{figure}

\end{document}